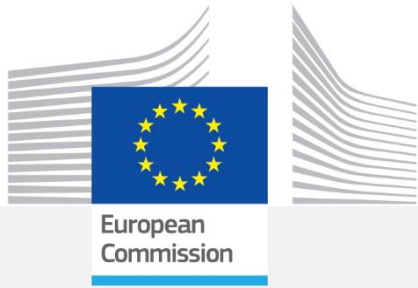

# Retrieval Augmented Generation Evaluation for Health Documents

*An evaluation and usability study*

Ceresa, M; Bertolini, L., Comte, V.; Spadaro N.; Raffael, B.; Toussaint, B.; Consoli, S.; Muñoz Piñeiro A.; Patak, A.; Querci M.; Wiesenthal T.

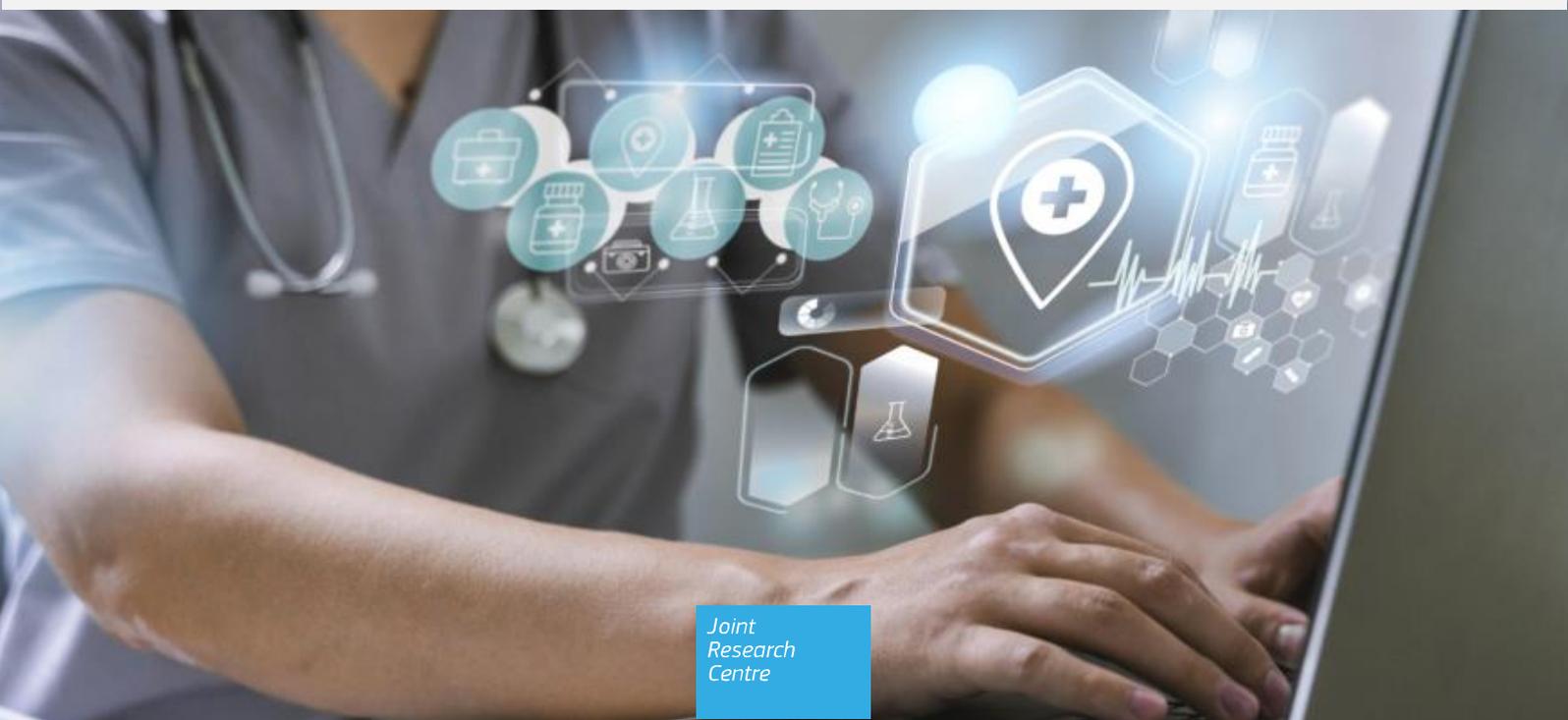

**Contact information**
Name: Mario Ceresa
Address: Via E. Fermi 2749, 1 – 21027 Ispra, Italy
Email: Mario.CERESA@ec.europa.eu
Tel.: +39 0332 78 9634

**EU Science Hub**
https://joint-research-centre.ec.europa.eu

Luxembourg: Publications Office of the European Union

© European Union

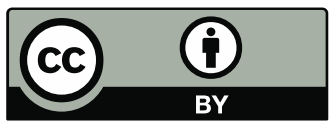

How to cite this report: Ceresa, M; Bertolini, L., Comte, V.; Spadaro N.; Raffael, B.; Toussaint, B.; Consoli, S.; Muñoz Piñeiro A.; Patak, A.; Querci M.; Wiesenthal T. *Retrieval Augmented Generation Evaluation for Health documents*, Publications Office of the European Union, Luxembourg, JRC138904.

**Table of Contents**

# Table of Figures

# 1 Abstract

Safe and trustworthy use of Large Language Models (LLM) in the processing of healthcare documents and scientific papers could substantially help clinicians, scientists and policymakers in overcoming information overload and focusing on the most relevant information at a given moment. Retrieval Augmented Generation (RAG) is a promising method to leverage the potential of LLMs while enhancing the accuracy of their outcomes. This report assesses the potentials and shortcomings of such approaches in the automatic knowledge synthesis of different types of documents in the health domain. To this end, it describes: (1) an internally developed proof of concept pipeline that employs state-of-the-art practices to deliver safe and trustable analysis for healthcare documents and scientific papers called RAGEv (Retrieval Augmented Generation Evaluation); (2) a set of evaluation tools for LLM-based document retrieval and generation; (3) a benchmark dataset to verify the accuracy and veracity of the results called RAGEv-Bench. It concludes that careful implementations of RAG techniques could minimize most of the common problems in the use of LLMs for document processing in the health domain, obtaining very high scores both on short yes/no answers and long answers. There is a high potential for incorporating it into the day-to-day work of policy support tasks, but additional efforts are required to obtain a consistent and trustworthy tool.

## 1.1 Authors' contributions and conflicts of interest

This report was a group effort from many people from JRC.F7: Mario Ceresa ideated the study together with Amalia Muñoz, programmed the initial implementation of the reference pipeline, and wrote most of chapters 2, 3 and 4. Amalia Munoz co-ideated the study and oversaw the three use cases of Virtual Human Twins, Horizon Research Projects and Bacteriophages. She also prepared the questions and answers for the Virtual Human Twins use case and contributed to the writing of the corresponding sections. Lorenzo Bertolini prepared the experiments on PubMedQA, the SHy pipeline and contributed to the writing of the methods and results chapters. Together with Barbara Raffael, he conceived the use case on Horizon Europe projects. Barbara Raffael prepared the questions for the Horizon Europe projects and contributed to the writing of the corresponding sections. Nicholas Spadaro prepared the frontend and backend part of the reference pipeline and contributed to the running of the experiments. Valentin Comte and Brigitte Toussaint prepared the use case on Antimicrobial Resistance and contributed to the writing of the corresponding sections. Sergio Consoli wrote the part of Colbert evaluation and reviewed the document. Tobias Wiesenthal advised and coordinated the study and reviewed the manuscript together with Maddalena Querci and Alex Patak.

All authors declare that they have no personal interest, in particular a family or financial interest, or representation of any other interest of third parties, which may lead to an actual or potential conflict of interest with respect to the results presented in this report.

## 1.2 Acknowledgements

We would like to thank Mr. Riccardo Medinas and Prof. Diego Reforgiato Recupero from the University of Cagliari for providing the implementation and results of the analysis of the ColBERT pipeline. We are also grateful to colleagues from the GPT@JRC team for providing an infrastructure to run Large Language Models experiments at scale in a safe environment. We would also like to thank colleagues from SANTE.R.4 and RTD.G.6 for the fruitful discussions and all colleagues from JRC.F who tested the system and provided useful feedback.

## 2 Introduction and related work

Over recent years, there has been an unprecedented surge in the volume of scientific publications and policy reports across various fields. Just between 2014 and 2024, 10,036,232 documents and 2,027,823 preprints have been published in Scopus database in the health domain.[1] This exponential growth [1], while bringing an enormous richness of accessible knowledge, also poses challenges for researchers, policymakers, and practitioners in view of keeping updated on contemporary research outcomes. The sheer volume of information makes it increasingly difficult to stay at the forefront of the latest developments, identify emerging trends, and pinpoint gaps in existing research. Traditional methods of literature review, reliant on manual curation and analysis, are no longer sufficient to manage the overflow of information.

Large language models (LLMs) have emerged as a powerful tool in natural language processing, with potential applications across numerous domains [2]. In the healthcare sector, these models offer promising opportunities to process, analyze, and generate insights from vast amounts of documentation [3]. LLMs, with their ability to understand context, extract relevant information, and generate human-like text, present a novel approach to addressing these challenges of information overload, but come with their own set of problems and limitations.

### 2.1 Objectives and key results

The purpose of this report is threefold: first, to conduct an exhaustive investigation into the current best practices for the safe and trustworthy utilization of LLM in the processing of healthcare data and scientific papers. Second, it aims to introduce an internally developed proof of concept pipeline that employs these practices to deliver safe and trustable processing for scientific papers in the health domain called RAGEv[4] (Retrieval Augmented Generation Evaluation), to assess the potential and drawbacks of the methods for different types of documents in the health domain. Furthermore, it aims to disseminate a benchmark dataset to verify the accuracy and veracity of the results called RAGEv-Bench[5]. Finally, it draws conclusions and provides recommendations on the use of the method and its actual way of implementation.

Our main conclusions are that careful implementations of RAG techniques could minimize most of the common problems in the use of LLMs for document processing, obtaining very high scores both on short yes/no answers and long answers. There is a high potential for incorporating it into the day-to-day work of policy support tasks, but additional efforts are required to obtain a consistent and trustworthy tool.

### 2.2 Structure of the document

Chapter 3 introduces the state of the art of the technologies used in this report, the context of the problem and the possible solutions. Chapter 4 explains in detail the methodology, with special attention to the Retrieval Augmented Generation technique, the different datasets used in both the automatic evaluation and the use cases usability checks and the design of the experiment. Chapter 5 presents the results and discusses the impact of the chosen methodology on both the automatic and manual evaluation. Chapter 6 contains the conclusions, the lesson learnt and some advice and future working directions. Additional online materials are on the DigLife project site[2]. Interested readers will find online all the materials and code to replicate the study, including the full use case questions here.

---

[1] accessed 1/7/2024

---

[2] https://diglife.ec.europa.eu

# 3 Background on the use of LLMs for health documents

## 3.1 Large Language Models

Language Models (LMs) are neural networks (pre)trained on massive amounts of text [4], designed to encode or produce sequences of words. In recent years, they have shown impressive results on a plethora of tasks [2] and they have become the model of choice in the natural language processing (NLP).

Current LMs have evolved from statistical approaches that use deterministic methods to represent language as a probability distribution. For example, unigram LM modelled language as the probability of encountering a single word, like "dog", given a large set of training data, such as Wikipedia. Current state-of-the-art LMs are based on the Transformer neural architecture [5], the main component of which is the self-attention mechanism [6].

In simple terms, the attention mechanism allows these models to focus on different parts of a sentence or document, much like how humans pay attention to different words when understanding language. For instance, in the sentence "The cat sat on the mat," the attention mechanism might assign higher weights to "cat" and "mat" when determining where the cat is located. This mechanism enables LMs to process and understand context more effectively by giving more importance to relevant words while processing text.

Researchers have found that scaling up these models can lead to performance improvements. When the parameter scale exceeds a certain level, these language models show some special emergent abilities that are not present in small-scale language models [7]. Those emerging capabilities allow an LLM to solve even tasks they were not specifically trained for, given that the user writes a prompt with a description of the novel tasks and provides some examples. These approach is called In Context Learning (ICL). To distinguish the difference in parameter scale, the research community has coined the term "large language models" (LLMs) for the LMs of significant size.

The research on LLMs has been largely advanced by industry in the last few years, due to the need of huge, concentrated investment in hardware and talent required. A remarkable milestone was the launch of ChatGPT by OpenAI, which has attracted widespread attention from society.

Among the many tools that currently use LLMs to process documents, ScopusAI[3] is a new functionality available in Scopus, a repository database of scientific publications. ScopusAI applies advanced technologies in the form "Question&Answer" to extract information from the publications. The user poses a specific question for which the system provides a comprehensive answer that includes references for each of the statements. In addition, it allows the extraction, in the form of an Excel table, of all relevant information related to the publication encoded in the metafile. Furthermore, the system creates a concept map corresponding to the answer provided, and it proposes three consequent prompts to dive into more specific details of different aspects of the answer. Additionally, many online and open source tools are available to implement systems for the use of LLMs to process documents such as LangChain[4], DiFy[5], Perplexity AI[6], Microsoft Copilot[7], and many more. All of those systems unfortunately suffer from the problems described in the next sections.

---
[3] https://www.elsevier.com/products/scopus/scopus-ai

## 3.2 Specific challenges for health documents

LLMs trained on public data may inadvertently expose private information if not properly managed [8], [9], [10], and this needs to be taken into account when using AI technologies to process health documents that are not in public domain (like open access scientific papers) or that contain sensitive personal information.

Furthermore, LLMs can perpetuate or amplify existing biases in medical data and practices reported in scientific studies, potentially leading to less accurate or biased results for underrepresented groups [11]. This raises concerns about equitable healthcare delivery and the potential for AI to exacerbate existing disparities in medical outcomes.

The lack of explainability and interpretability in LLMs [12] is particularly problematic for the use of retrieving information from health-related publications. On the one hand the "black box" nature of these models makes it difficult to understand how they arrive at conclusions, which can hinder their adoption by healthcare professionals who need to justify their decisions. This lack of transparency also complicates efforts to ensure that LLM outputs align with current medical best practices and guidelines. On the other hand, and despite their limitations, LLMs could redefine interpretability itself by being used to audit their own responses [13].

Factuality and trustworthiness are critical issues when applying LLMs to health documents and scientific papers. These models can generate plausible-sounding but incorrect information, often referred to as "hallucinations" [14]. But, in our case, even small inaccuracies can have serious consequences, and verifying the accuracy of LLM outputs in complex medical contexts is both challenging and resource intensive.

Toxicity and the generation of inappropriate content are also concerns that must be addressed [15], especially given the sensitive nature of medical information retrieval. Regulatory compliance adds another layer of complexity, since the development and use of AI in healthcare is subject to strict and evolving regulations such as the General Data Protection Regulation (GDPR)[8], the AI Act[9] and the Medical Devices Regulation (MDR)[10]. Ensuring compliance while leveraging the capabilities of LLMs requires careful navigation of legal and ethical frameworks.

Data quality and consistency pose significant challenges, as medical data in publications is often inconsistent, incomplete, or recorded in non-standard formats. LLMs may struggle with these inconsistencies, leading to unreliable outputs. Moreover, the deep domain knowledge required in medical contexts may be beyond the capabilities of general-purpose LLMs, leading to potential misinterpretations of nuanced medical information.

---

4 https://www.langchain.com/
5 Dify.AI · The Innovation Engine for Generative AI Applications
6 https://www.perplexity.ai/
7 https://copilot.microsoft.com/
8 https://eur-lex.europa.eu/eli/reg/2016/679/oj
9 https://digital-strategy.ec.europa.eu/en/policies/regulatory-framework-ai
10 https://eur-lex.europa.eu/legal-content/EN/TXT/?uri=CELEX%3A32017R0745

There is also a risk of over-reliance on LLM outputs, potentially leading to weakening critical thinking and automation bias where users might overlook their own expertise or (patient, organ, disease, specialty)-specific factors [16]. This raises broader ethical questions about decision-making, patient autonomy, and the role of AI in health at large.

Addressing these limitations requires implementing robust safeguards, continuous monitoring, and maintaining human oversight. Developing health-specific LLMs trained on high-quality, diverse, and multimodal data, and adhering to strict ethical and regulatory guidelines could mitigate some of these concerns. However, it is crucial to approach the use of LLMs in health document analysis with caution, ensuring that their use enhances rather than compromises the quality and equity of patient care.

### 3.3  Proposed solutions and approaches

We identified in the previous section several critical aspects for the use of LLMs in the processing, transforming and distillation of healthcare documents, namely: leakage of personal information, bias and fairness, lack of explainability or interpretability, forgetfulness and limited factuality and toxicity. Let's now summarize the main strategies present in the literature to mitigate those risks and thus unlock the full exploitation of this powerful AI tool.

To mitigate the risk of personal data leakage, robust de-identification techniques should be employed, along with federated learning approaches that keep sensitive data localized [17]. Differential privacy methods can also be applied to add noise to the data, further protecting individual privacy [18], [19]. Regular audits of model outputs should be conducted to detect and prevent potential information leaks.

To address bias and fairness issues, it is fundamental to diversify training data sources and employ bias detection and mitigation algorithms [20]. Regular testing of models for disparate outcomes across different demographic groups is essential, as it is the implementation of fairness constraints in model objectives.

The lack of explainability and interpretability in LLMs poses challenges in healthcare settings where transparency is critical. While in classical models locally interpretable machine learning techniques such as LIME[11] and SHAP[12] can be utilized to provide insights into model decision-making processes [21], [22], [23], these techniques do not work well for LLMs. Developing domain-specific explanation methods such as [12], using in-domain knowledge augmentation, and combining LLMs with rule-based systems can further enhance transparency. Additionally, providing confidence scores and uncertainty estimates can support healthcare professionals to better understand and trust the model outputs.

Forgetfulness and limited factuality are significant concerns when using LLMs for healthcare document processing. Implementing retrieval-augmented generation techniques can help addressing these issues by allowing the models to access and incorporate up-to-date information [24]. Regularly updating model knowledge bases and using fact-checking mechanisms are also crucial. Furthermore, combining LLMs with structured knowledge graphs can enhance their ability to provide accurate and current information.

---

[11] https://github.com/marcotcr/lime
[12] https://shap.readthedocs.io/en/latest/

Toxicity in LLM outputs can be particularly problematic in healthcare contexts. To mitigate this issue, models should be fine-tuned on curated, non-toxic datasets specific to healthcare, or undergo unlearning [25] of the unwanted content. Additional strategies could be implementing content filtering and moderation systems, along with toxicity detection algorithms, which can help preventing the generation of harmful or inappropriate content [26].

We believe that - by addressing the abovementioned critical aspects the use of LLMs in healthcare document processing can be made safer, more reliable and effective. However, it is important to note that this field is rapidly evolving, and ongoing research is necessary to continuously improve these systems and address emerging challenges.

# 4 Methods for trustworthy use of LLMs to health documents

## 4.1 Strategies for improving the accuracy of the LLM-generated answer

Finetuning, prompt engineering and RAG, are three distinct approaches that can be employed to improve the performance of language models in terms of accuracy, each with its own characteristics and use cases.

**Finetuning** is the process of further training a pre-trained model on a specific dataset or task. This approach adapts the model's weights to perform better on the target task and can significantly improve performance for specialized applications, such as the ones in the health domain, where a general LLM might not have already encountered the relevant vocabulary before. This is however a computationally intensive operation, especially for LMs of large size.

**Prompt engineering** is a technique that focuses on crafting effective prompts to guide model behavior. Unlike finetuning, prompt engineering does not modify the model itself but instead optimizes the input to achieve desired outputs. This method can be used to improve performance without additional training and relies on understanding model behavior and crafting precise instructions. In LLMs, prompt engineering allows the use of In Context Learning (ICL) which, as discussed in Section 3.1, is an emerging property of models with a large number of parameters that enables them to generalize and apply learned knowledge to new tasks without the need for explicit retraining or fine-tuning. By designing prompts in specific ways, users can guide the model to perform tasks in a zero-shot or few-shot manner, leveraging the model's ability to infer the desired output based on the context provided within the prompt itself. [27].

**Retrieval Augmented Generation** combines a pre-trained language model with a retrieval system. This method retrieves relevant information from a well-defined external knowledge base to inform the generation process, allowing for up-to-date and factual responses without the need to retrain the entire model. RAG is particularly useful for question answering and tasks that require current or domain-specific knowledge.

The key differences between these approaches lie in how they modify the model, use external data, their flexibility, and resource requirements. Finetuning is the only method that actually modifies the model's weights, while RAG and prompt engineering leave the model's weights unchanged. Additionally, RAG uses external data during inference, finetuning incorporates it during training, and prompt engineering does not necessarily use external data at all.

In terms of flexibility, prompt engineering is the most adaptable, as it can be quickly adjusted for different tasks, but its more powerful version, ICL, is only available on very large models, having a sufficiently extensive context length, and does not always work consistently as expected. RAG allows for easy knowledge updates without retraining, while finetuning is the least flexible but potentially most powerful for specific tasks. It might also be the only possibility for tasks that involve a new vocabulary, not present in the original pretraining of the LLM. Regarding resource requirements, finetuning typically demands the most computational power, followed by RAG, with prompt engineering being the least resource-intensive.

Each of these methods comes with its strengths and limitations, and is suited to different scenarios, depending on the specific requirements of the task at hand, available resources, and the desired balance between performance and flexibility. For additional details on whether it is better to use a generalist LLM and prompt it with examples of the desired task (thus using ICL) or to finetune a smaller and specialized model, we refer the interested reader to [28], [29], [30], [31]. For additional details on how to implement RAG, excellent reviews are [32], [33].

## 4.2 Taxonomy of a classical RAG system

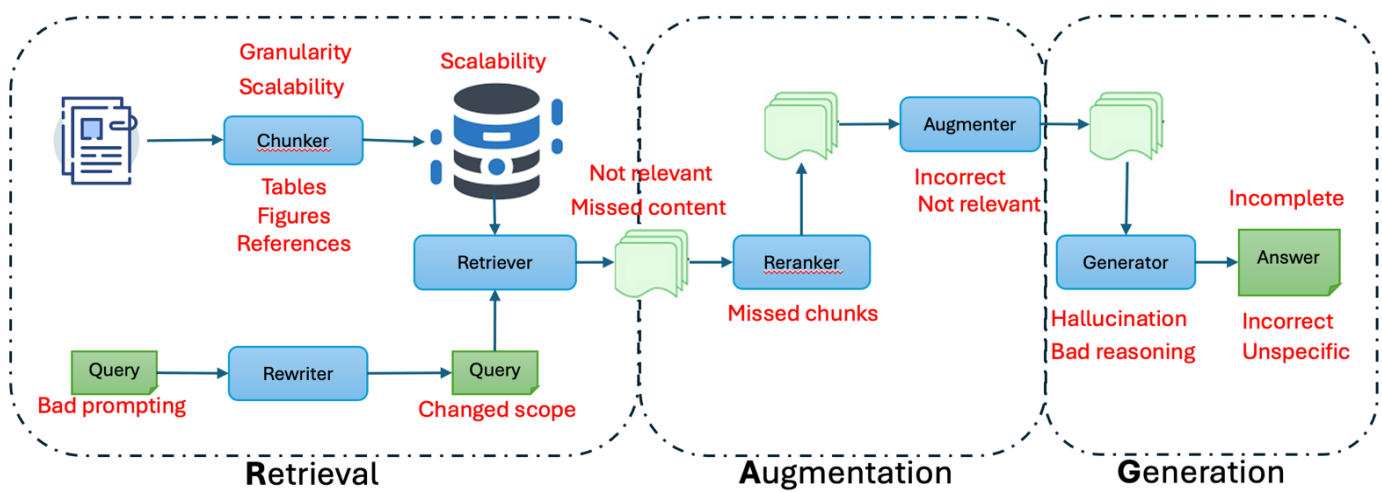

*Figure 1 - Main elements in a RAG pipeline and most common problems*

As shown in Figure 1, traditional RAG includes three main steps: indexing, retrieval, and generation. Indexing starts with the cleaning and extraction of raw data in diverse formats like PDF, HTML, Word, and Markdown, which is then converted into a uniform plain text format. To accommodate the context limitations of language models, text is segmented into smaller, digestible chunks. Chunks are then encoded into vector representations using an embedding model and stored in a vector database. This step is crucial for enabling efficient similarity searches in the subsequent retrieval phase.

### 4.2.1 Indexing

The most common chunking strategy is to split the document into chunks on a fixed number of tokens (e.g., 100, 256, 512) [34]. Larger chunks can capture more context, but they also generate more noise, requiring longer processing time and higher costs. While smaller chunks may not fully convey the necessary context, they introduce less noise. However, chunking might lead to truncation within sentences, and uncertainty on the best granularity of the embedding. Some strategies seek to optimize that with recursive splits and sliding window methods, enabling layered retrieval and merging globally related information across multiple retrieval processes [35]. Nevertheless, these approaches still cannot strike a balance between semantic completeness and context length. Therefore, methods like Small2Big have been proposed, where sentences (small) are used as the retrieval unit, and the preceding and following sentences are provided as (big) context to LLMs. Additional methods are establishing a hierarchical index or using knowledge graphs [36].

### 4.2.2 Embeddings

Text embeddings are dense vector representations of text data. In this context, words and documents with similar meanings are represented by similar vectors in a high-dimensional vector space [37] [38]. Thus, embeddings encode semantic information and serve as a foundation for various downstream applications, including retrieval, reranking, classification, clustering, and semantic textual similarity tasks. Additionally, the embedding-based retriever plays a crucial role in retrieval-augmented generation [24], which allows LLMs to access the most up-to-date external or proprietary knowledge without modifying the model parameters [39], [40], [41], [42].

Word2Vec [43] was the first neural network-based approach to predict surrounding words for a target word in each context. It learns to encode word semantics into vectors. GloVe [44] (Global Vectors for Word Representation) leverages global statistics to create embeddings. It captures co-occurrence patterns between words in large text corpora.

BERT [38] (Bidirectional Encoder Representations from Transformers), a transformer-based model, generates contextualized word embeddings by considering both left and right context. This is one of the most widely used for various NLP tasks. Furthermore, SBERT [45] (Sentence BERT) is specifically designed for sentence similarity tasks. While BERT embeddings can be used for sentences, SBERT embeddings are more effective for capturing sentence semantic similarity.

Decoder-only LLMs [7] were believed to underperform bidirectional models on general-purpose embedding tasks due to unidirectional attention that limits the representation learning capability, along with the scaling of LLMs which leads to very high-dimension embeddings, that may in turn suffer from the curse of dimensionality.

Nevertheless, the latest text-embedding-3-large from OpenAI obtained a MTEB score of 64.59[13] in 2024, opening the way to practical use of LLMs as embedders. Furthermore, E5-Mistral [46] obtained a METB score of 66.63, applying contrastive learning with task-specific instructions on Mistral 7B [47]. It outperforms the state-of-the-art bidirectional models on comprehensive embedding benchmarks [48] by utilizing a massive amount of synthetic data from the proprietary GPT-4 model. LLM2Vec [49] (METB score: 65.01) tries to build the embedding model from LLMs while only using public available data, but it is still worse in performance than E5-Mistral.

---
[13] https://openai.com/index/new-embedding-models-and-api-updates/

Given the notable success of E5-Mistral, SFR-Embedding-Mistral [50] (METB: 67.56) further fine-tunes it on the blend of non-retrieval and retrieval datasets for improved accuracy on both tasks. The latest embedding model is NV-Embed from NVIDIA, which is very closely related to the SFR approach but uses only public available data and is not dependent on other embeddings models [51].

At the time of the writing of this report, this was the latest ranking of embedding models on the MTEB benchmark [48]:

| Model | Model Size (Million Parameters) | Memory Usage (GB, fp32) | Embedding Dimensions | Max Tokens | Average (56 datasets) | Classification Average (12 datasets) | Clustering Average (11 datasets) | PairClassification Average (3 datasets) | Reranking Average (4 datasets) |
|---|---|---|---|---|---|---|---|---|---|
| SFR-Embedding-2_R | 7111 | 26.49 | 4096 | 32768 | 70.31 | 89.05 | 56.17 | 88.07 | 60.14 |
| gte-Qwen2-7B-instruct | 7613 | 28.36 | 3584 | 131072 | 70.24 | 86.58 | 56.92 | 85.79 | 61.42 |
| neural-embedding-v1 | | | | | 69.94 | 87.91 | 54.32 | 87.68 | 61.49 |
| NV-Embed-v1 | 7851 | 29.25 | 4096 | 32768 | 69.32 | 87.35 | 52.8 | 86.91 | 60.54 |
| voyage-large-2-instruct | | | 1024 | 16000 | 68.28 | 81.49 | 53.35 | 89.24 | 60.09 |
| Linq-Embed-Mistral | 7111 | 26.49 | 4096 | 32768 | 68.17 | 80.2 | 51.42 | 88.35 | 60.29 |
| SFR-Embedding-Mistral | 7111 | 26.49 | 4096 | 32768 | 67.56 | 78.33 | 51.67 | 88.54 | 60.64 |
| gte-Qwen1.5-7B-instruct | 7099 | 26.45 | 4096 | 32768 | 67.34 | 79.6 | 55.83 | 87.38 | 60.13 |
| gte-Qwen2-1.5B-instruct | 1776 | 6.62 | 4096 | 131072 | 67.16 | 82.47 | 48.75 | 87.51 | 59.98 |
| voyage-lite-02-instruct | 1220 | 4.54 | 1024 | 4000 | 67.13 | 79.25 | 52.42 | 86.87 | 58.24 |

*Figure 2 First ten embedding models from MTEB leaderboard (retrieved from HuggingFace, 07/07/24)*

### 4.2.3 Retrieval

Upon receipt of a user query, the RAG system employs the same encoding model utilized during the indexing phase to transform the query into a vector representation. It then computes the similarity scores between the query vector and the vector of chunks within the indexed corpus. The system prioritizes and retrieves the top K chunks that demonstrate the greatest similarity to the query. These chunks are subsequently used as the expanded context in prompt.

The retrieval phase often struggles with precision and recall, leading to the selection of misaligned or irrelevant chunks, and the missing of crucial information.

### 4.2.4 Generation

The posed query and selected documents are synthesized into a coherent prompt to which a large language model is tasked with formulating a response. The model's approach to answering may vary depending on task-specific criteria, allowing it to either draw upon its inherent parametric knowledge, or restrict its responses to the information contained within the provided documents. In cases of ongoing dialogues, any existing conversational history can be integrated into the prompt, enabling the model to engage in multi-turn dialogue interactions effectively.

In generating responses, the model may face the issue of hallucination, where it produces content not supported by the retrieved context. This phase can also suffer from irrelevance, toxicity, or bias in the outputs, detracting from the quality and reliability of the responses. The process may also encounter redundancy when similar information is retrieved from multiple sources, leading to repetitive responses. Determining the significance and relevance of various passages and ensuring stylistic and tonal consistency add further complexity, as it will become clear from the next section.

## 4.3 Reference implementation (RAGEv)

As we discussed in the previous section, traditional RAG faces two important challenges. The first one is in the retrieval phase, where it often struggles with precision and recall, leading to the selection of misaligned or irrelevant chunks, and the missing of crucial information. The second one is the generation phase, where it might produce content which is not supported by the retrieved contexts, or that also be irrelevant, toxic or redundant.

A reference implementation of several pipelines that use LLMs to analyse scientific publications and policy documents within the health domain was preparedperformed, as illustrated in the following. Such pipelines go collectively under the name of RAGEv (Retrieval Augmented Generation Evaluation) and are used in all the subsequent experiments of this report. The RAGEv[14] system is to be considered a proof of concept and a working tool that, while extremely useful for the preparation of this report, is not scalable enough to be used as a full Information System. Interested readers are welcome to use it carefully to reproduce the results of this report, or to benchmark their own use cases, but please consider its purpose to serve as a proof-of-concept rather than as operational tool.

Selected functionalities will be embedded inside the AI for Health space of GPT@JRC, integrating them into one single corporate tool. GPT@JRC is a platform for scientific and non-scientific staff of the EC (European Commission) to securely explore the use of pre-trained Large Language Models. It offers secure access to a wide variety of text Generative AI models to assist with written office tasks and support scientific work. User prompts and the generated responses are not shared with third parties. GPT@JRC allows the users to interact and work with different Large Language Models, such as the commercial OpenAI GPT family models (currently: GPT 3.5, GPT 3.5 large, GPT 4, GPT 4 32k, GPT4 Turbo), and some powerful open source models (such as: Nous Hermes Mixtral, Llama 3 70B chat, MistralOrca 7b, Zephyr beta 7b[15]). Our proposed system, RAGEv, could be used to further evaluate for factuality and trustworthiness the RAG functionalities supported by GPT@JRC, including the possibility to use PDF and Word documents in chats, and the integration of GPT@JRC with internal bodies of knowledge and tools.

---

[14] https://vast.jrc.ec.europa.eu
[15] as accessed on 1/7/2024

### 4.3.1 Retrieval pipelines

Several pipelines were developed to check the performance of different approaches on retrieval results:

*Table 1 - Retrieval pipelines tested*

| Pipeline | Description |
|---|---|
| Vanilla | A vanilla LLM is a basic, unmodified version of a large language model. It is trained on a large corpus of text and can generate human-like text based on the input it is given. However, it does not have any additional features or modifications that might be used to improve its performance or adapt it to specific tasks. |
| Vector Search | Vector Search uses vectors to represent and efficiently search through complex, unstructured data. It represents data points as vectors in a high-dimensional space, where each dimension defines a specific attribute or feature. The search compares the similarity of the query vector to the possible vector paths that traverse all of the dimensions. |
| Full-text Search | Full-text search refers to techniques for searching a single computer-stored document or a collection in a full-text database. In a full-text search, a search engine examines all of the words in every stored document as it tries to match search criteria (for example, text specified by a user). It is distinguished from searches based on metadata or on parts of the original texts represented in databases. |
| Hybrid Search with Reranking | Hybrid search is a combination of full-text and vector queries that execute against a search index that contains both searchable plain text content and generated embeddings. It combines results from both full-text and vector queries, which use different ranking functions. A Reciprocal Rank Fusion (RRF) algorithm merges the results. The query response provides just one result set, using RRF to pick the most relevant matches from each query. |
| SHy | SHy (for **S**ingle **Hy**brid) implements the hybrid search with reranking pipeline, with the difference that each document in the collection is treated as a single collection. The design and implementation of this pipeline was motivated by the necessity of acquiring more space and horizontal knowledge from all the provided documents, so to answer broader and summarisation-oriented questions that users might have about the whole collection. |
| ColBERTv2 | ColBERT is based on Late Interaction, introduced in [52], which encodes each passage into fine-grained token-level embeddings and finds passages that contextually match a query using scalable vector-similarity operators. ColBERTv2 [53] is a follow-up to ColBERT achieving 6-10x storage reduction while maintaining performance. It uses compression techniques and denoised supervision. |

### 4.3.2 System architecture

The system is structured as in Figure 3, in a hybrid cloud architecture, with part of the system running on premises in the JRC datacentre, and part in Azure cloud. The former include the landing page of the DigLife project, along with the frontend of our RAG-Ev reference pipeline. The latter is the backend system exposing the APIs used to manage the knowledge bases and interact with them, leveraging large language model and dedicated pipelines.

The frontend has been developed in NextJS[16] and is deployed on internal JRC Openshift, under EULogin[17], and takes care of user authentication and communication with the backend.

The backend system is deployed on Azure Managed Kubernetes instances[18]. This system allows CRUD (Create, Read, Update, and Delete) operations on collections of documents, their embeddings in vector DBs such as Milvus[19] and to control the various retrieval pipelines. A set of APIs is provided to control all functionalities via python scripts.

The backend system also manages communications with the JRC's Big Data Platform (BDAP)[20] for model training and finetuning, and with GPT@JRC for using large language models in a safe environment.

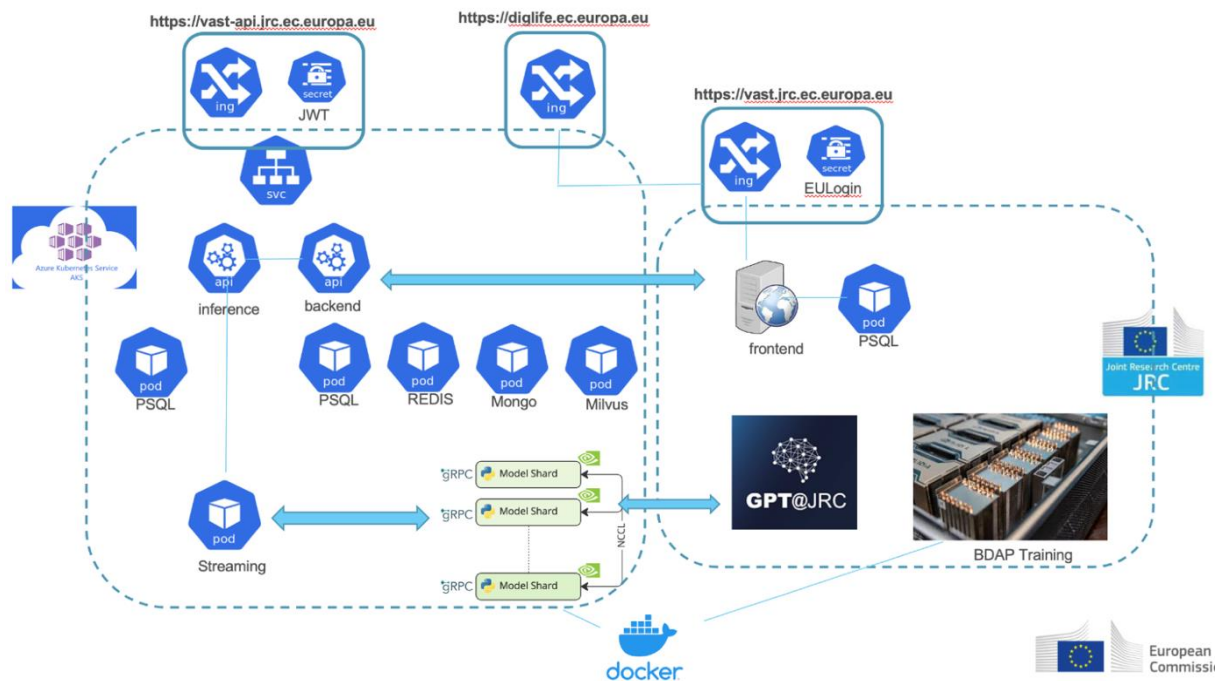

*Figure 3 - Main architecture of the system. We use a hybrid cloud approach, where the model training is done on premises on BDAP servers. Most of the experiments use the LLMs served by the GPT@JRC infrastructure.*

## 4.4 Datasets for benchmarking and usability checks (RAGEv-Bench)

To systematically evaluate our system and its configurations, we used a two-fold approach. First, we ran an automatic evaluation based on public datasets and on classical automatic metrics. This part of the evaluation, called in subsequent sections the "Automatic performance evaluation" (APE), sets the baseline and allows comparison with existing methods and implementation on public and well-known datasets. This step has also been used to define many of the hyperparameters of the RAG pipelines that are then used in the second evaluation. As a second evaluation, we are interested in the usability of these pipelines in the day-to-day work of colleagues which are not experts in AI and work on policy support files. To this end, we built three hand-crafted datasets - referred in the subsequent sections as "Horizon Research" (HR), "Virtual Human Twin" (VHT) and "Bacteriophages" (AMR) - that cover three topics of the current work of the Unit JRC.F7. The goal was to assess the system on documents and questions that users across the EC could present to the system and learn from the feedback of colleagues with deep domain knowledge on how to facilitate the integration of LLMs with their workflow. This assessment allowed further improvement of RAGEv.

---

[16] Next.js by Vercel - The React Framework (nextjs.org)
[17] https://webgate.ec.europa.eu/cas/login
[18] https://azure.microsoft.com/en-us/products/kubernetes-service
[19] https://milvus.io/
[20] https://jeodpp.jrc.ec.europa.eu/bdap/

All together, these four datasets are released under the name of RAGEv-Bench for interested readers to replicate our results and advance research. The datasets will be available in the JRC Data catalogue in the weeks after the publication of this report.

### 4.4.1 Automatic performance evaluation on PubMedQA (APE)

The APE dataset is a subset of PubMedQA[21], containing instances of data used for biomedical research Question Answering tasks and built upon a large collection of PubMed papers. It contains questions that are built, either automatically or manually, from the title of each work. Doing so, the dataset allows for each query to be answered in a yes/no/maybe fashion. To help the model answer the question, the dataset contains a set of 3-4 short textual chunks for each query, extracted by the extended abstract of each paper. Altogether, the PubMedQA dataset contains up to two 200 k instances, each containing a PubMed ID, a question, a context (i.e., the short textual snippets), a short answer (yes/no/maybe), and a longer answer. Since the scope of our assessment study is to evaluate RAGEv on full documents, we focused on a subset of 100 documents, extracted using the PubMed ID. This number was in the end reduced to 88, as we considered only the ones in public access to ensure reproducibility of this study.

### 4.4.2 Horizon research (HR)

This collection focuses on the analysis of the scientific output of research projects. The EU is successfully funding research and innovation projects via its dedicated Framework programme, which generate an impressive amount of knowledge and data, accessible via CORDIS[22]. SOPLAS[23], a project on Macro and Microplastic in Agricultural Soil Systems, was chosen to test the performance of the RAGEv pipelines with unrelated documentation. SOPLAS reports 3 scientific papers and 2 additional documents, one of which with potentially side-tracking content (not related to microplastics, but on communication of results). For this, 10 questions ranging from very general questions, that is covering multiple documents, to very specific ones together with the ground truth answer, were prepared. The questions,

---

[21] https://pubmedqa.github.io/
[22] https://cordis.europa.eu/ - EU repository information about EU Research & Development projects
[23] SOPLAS, a project on Macro and Microplastic

based on the answer provided by the system, were adapted in the different phases on the test.

### 4.4.3 Virtual Human Twins (VHT)

This collection focuses on a relatively new scientific health topic, having non-homogeneous information coverage, that is redundant information related to barriers, but much less on specific applications. A virtual human twin (VHT) is a digital representation of different levels of human anatomy (e.g. cells, tissues, organs, or organ systems) for modelling the state of human health or disease state[24]. RAGEv was tested for its performance when investigating the state of the art of research in the topic, as well as its drivers and barriers. Of the 2,000 documents identified in Scopus Database, responding to the specific search string[25], 10 documents related to the subtopic "Main constrains for the implementation of digital twin in health care" were selected and uploaded as a collection for further evaluation with RAGEv. For this, 18 questions ranging from very general questions, that is covering multiple documents, to very specific ones, together with the ground truth answer, were prepared.

### 4.4.4 Bacteriophages (AMR)

This collection covers a topic well stablished and mature in the scientific health domain. Phage therapy consists of administering viruses, called phages or bacteriophages, as medicine to kill pathogenic bacteria in case of severe bacterial infection, especially when other antimicrobial drugs are not efficient. They are very specific and not harmful to humans since they only infect bacteria. They are used in human patients but also in livestock and companion animals. However, phage therapeutic products are not yet authorised on the market. There are currently no sufficient evidence-based clinical protocols, no complete dedicated regulatory framework and it is still necessary to investigate the impact of phages on the environment before implementing phage therapy on a large scale.

In this context, RAGEv was tested in its capacity to help gathering information, highlighting key factors and knowledge gaps. Eight reviews were manually selected from the 15 reviews identified in the Scopus Database, responding to the specific search[26]; 8 were manually selected for testing RAGEv. For this selected set, 8 questions, addressing different types of information, such as numerical values, open questions, more specific questions, contained in a table, in a figure, or in section titles, together with the ground truth answer, were prepared.

## 4.5 Experiments

To ensure the efficacy and reliability of RAGEv, we run first a set of experiments on APE and then the manual usability tests on HR, VHT and AMR datasets.

---

[24] https://digital-strategy.ec.europa.eu/en/policies/virtual-human-twins
[25] "(("digital twin*") OR ("process twin*") OR ("data twin*")) AND (("healthcare") OR ("health care"))"
[26] 'resistance' AND 'phage' AND 'therapy' AND 'environment' AND 'one health'

### 4.5.1 Evaluation metrics

In all the cases where the answers of the system could be interpreted as a classification task (e.g., a binary yes/no response or an output that could be divided in classes) we used accuracy, precision, recall and F1 Score to present the results of the analysis.

For the experiments that assessed the generative abilities of the system with respect to a reference answer provided by a human evaluator, we make use the Rouge and BERT Scores. While the former are more lexicon oriented the latter is more focused on the semantic. In other words, while Rouge Scores compare the answer of a system to the desired one based on the content of their *actual* words, BERT Scores are more interested on how well two strings of text align on their overall semantic meaning.

Originally introduced to evaluated NLP models on summarisation, Rouge Scores are recall-based evaluation systems that search for the overlap between model's prediction and desired output at the n-gram level. For example, with n = 1, Rouge 1 will measure the overlap of single words between what the model has predicted, and the gold standard, following the equation:

$$Rouge - N = \frac{\sum_{gram_n \in S} Count_{match}(gram_n)}{\sum_{gram_n \in S} Count(gram_n)}$$

where *n* stands for the length of the n-gram $gram_n$, $Count_{match(gram_n)}$ is the maximum number of n-grams (a single word, in the case n = 1, or uni-gram) co-occurring in the gold standard answer and the answer produced by the system. While Rouge 1 and 2 respectively set n =1 and n = 2, Rouge L refers to the longest common subsequence shared between automatic and gold standard answers. The difference between Rouge L and LSum lays in the fact that Rouge LSum considers each sentence in a given answer independently, whereas Rouge L does not.

Given two textual strings A and B, BERT Scores [54] use a multi-step procedure to find a word-level alignment between A and B, and later measure a score for the pair, by weighting the semantic similarity between aligned words, using cosine similarity. To obtain such a score, authors have adapted precision, recall and F1 scores (i.e., BERT Score Precision, BERT Score Recall, BERT Score F1). In this work, we make use of all three subset of BERT Scores to evaluate generative answer (i.e., non-class-based) in each experiment.

### 4.5.2 Automatic performance evaluation (APE)

The reference dataset for the APE is a subset of PubMedQA, as described in Section 4.3.2. All system configurations in Figure 3 were evaluated using a Vanilla model and RAG pipeline, to answer medical questions. From the questions, we extracted both a yes-no short answer and a long, more detailed explanation. To measure the performance of the system, both standard metrics, such as accuracy and F1 scores, as well as generation-assessing metrics, namely Rouge Scores were adopted. More details can be found in previous Section 4.5.1

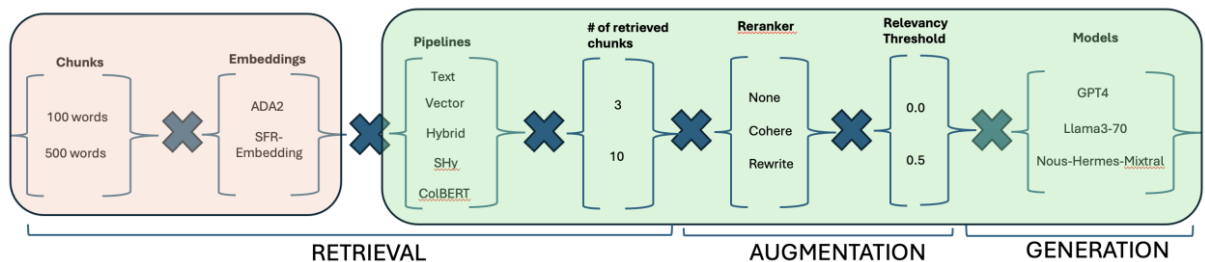
*Figure 4 - Design of experiments: we test both the Retrieval and Generation part of the system.*

The experiments are designed as multifactor with a structure of 2x2x5x2x3x2x3 as depicted in Figure 4. This amounts to a total of 720 experiments, for which we use the mnemonic codes CKw-EMB-PIP-#c-RER-RTH-MOD such as, for instance: 10-ADA-TEX-GPT, 50-SFR-SHY-NOU etc…

We additionally add a set of 3 experiments without the use of RAG, that is, interrogating directly the LLM without providing any context, to establish the baseline. We will identify those experiments as NORAG-{GPT, LLA, NOU}.

To analyse the results of our factorial experiment, we have used two approaches. First, a four-way Analysis of Variance (ANOVA) to examine the main effects of each factor and their interactions. In this case, we first verify that our data meets the ANOVA assumptions: normality of residuals, homogeneity of variances, and independence of observations. Then we use *pandas* and *penguin* statistical packages in *python* to calculate the main effects for each factor, two-way interactions, three-way interactions, and the four-way interaction.

For the interpretation of the results, we focus on the F-values and p-values for each main effect and interaction. Significant main effects indicate that the factor levels differ in their effect on the dependent variable, while significant interactions suggest that the effect of one factor depends on the levels of another factor. If significant main effects or interactions are found, we conduct post-hoc tests, such as Tukey's HSD, to determine which specific levels or combinations differ significantly. Additionally, we calculate effect sizes (e.g., partial eta-squared) to help understand the magnitude of significant effects.

Given the complexity of a four-way ANOVA, we have to ensure that we have an adequate sample size for each cell in the design and we want to conduct a power analysis to ensure the study has sufficient power to detect effects. If certain higher-order interactions are not significant or theoretically meaningful, we might consider simplifying the model.

As a second option, we used Linear mixed models (LMMs), as they offer a robust alternative to ANOVA for analysing factorial experiments, particularly when dealing with complex designs or unbalanced data. We will compare the two models:

- Human Score ~ Pipeline * Collection * Type of qQestion + BERTScore F1 + (Question id | 1))
- with a base one (i.e., Human Score ~ BERTScore F1 + (question | 1)).

Under the hypothesis that this is a fully crossed design where all "subjects" experience all 60 conditions. The model structure accounts for the factorial design in the fixed effects, similar to ANOVA, while also incorporating random effects to handle variation due to sampling or other sources of non-independence. LMMs offer several advantages over ANOVA, including better handling of unbalanced designs, the ability to model both categorical and continuous predictors, more flexible error structures, and improved handling of missing data.

### 4.5.3 Manual usability checks on HR, VHT and AMR datasets

The usability checks are performed to qualitatively evaluate the performance of the system from the domain expert point of view and provide feedback on how useful LLMs could be in the day-to-day work of colleague working on science for policy support. This evaluation is meant to provide some insights into the quality of the answers, as well as identify potential issues and hints to be considered when using similar tools and during further developments.

The usability checks are run on the three collections RP, VHT and AMR described previously respectively in Section 4.4.2, 4.4.3 and 4.4.4.

To prepare the questions, we initially planned to follow the approach in [55], resulting in the creation of four types of collections, detailed in Table 2 - types of collections.

*Table 2 - types of collections*

| TYPE | Description |
|---|---|
| Relevant | Collections containing only documents relevant to the topic |
| Some noise | Collections containing some documents that are not relevant or related to the topic |
| Noise only | Collections containing just noise, that is, none of the documents contain information relevant to the topic |
| Contrafactual | Collections containing contradicting statements |

However, RAGEv is currently conceived to work with user-defined collections of relevant documents, that is, the ones without noise or contradictory information. Future extensions of RAGEv, not based on user-uploaded collections, will demand testing other types of collections.

There is a rich body of literature to evaluate the usability of AI systems [56]. The questions tried to address different types of information such as numerical values, open questions, more specific questions; but also, information located in different parts of the document (tables or text). In addition, two questions were prepared that were included in more than one document to see how the systems would respond to that.

The questions chosen for each of the collections are classified by type according to the following classification:

*Table 3 – Question types*

| Type 0 | Type 1 | Type 2 | Type 3 | Type 4 | Type 5 | Type 6 |
|---|---|---|---|---|---|---|
| General/Summary | Yes/No | Numerical | Table | Focused | Figure | Subtitles |

Furthermore, to harmonise the evaluation of the answers, the quality of the answer was marked with the following scale:

*Table 4 - Scores*

| Score 0 | Score 1 | Score 2 | Score 3 | Score 4 | Score 5 |
|---|---|---|---|---|---|
| No answer or incorrect answer | The general topic is understood, but not the scope of the question. | The information is misinterpreted. | The answer is correct but is missing important information. | The answer is correct but missing minor information. | The answer is correct and complete. |

As previously stated, the full set of questions for each collection is detailed online as explained in Section 2.3.

Given the small sample number of the manual usability checks, we use descriptive statistic and plots for the analysis of the results.

# 5 Results and discussion

## 5.1 Result on the automatic performance evaluation (APE)

These are the results and analysis from the experiment to measure the performance of the different RAGEv pipelines on a single benchmark, namely PubMedQA, as described in Section 4.4.1 and 4.5.2. Figure 5 summarises at very high level the results of the 720 experiments, comparing the different RAGEv pipelines (i.e., Vector, Full-text, Hybrid with Rerank, SHy and ColBERTv2), and grouping by all the other factors and levels. Each pipeline (y-axis) was analysed with respect to its performance on short answers (Yes/No) using the accuracy, and with respect to its performance on the long open
answer with Rouge and BERT Score metrics.

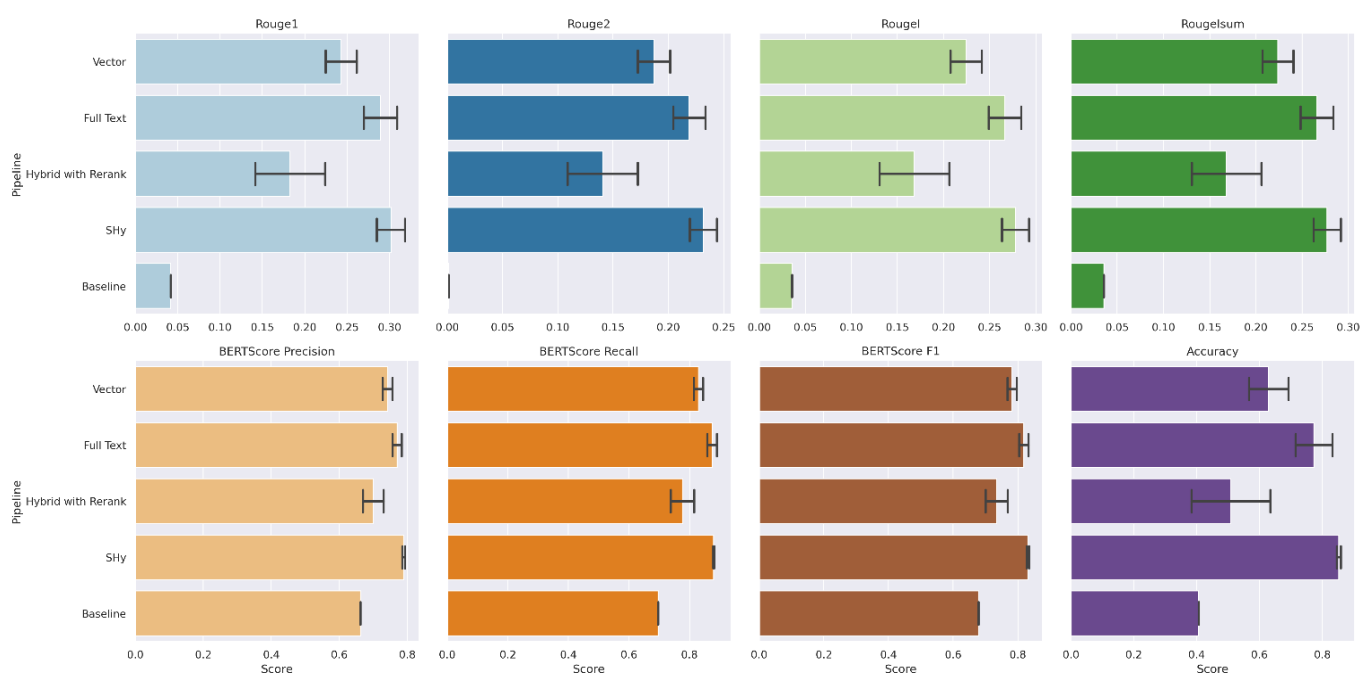

*Figure 5 – Pipeline comparison across several metrics.*

The Figure above strongly suggests that there is a marked improvement in the performance of the LLMs when adding a RAG component (baseline vs all) and that the variations within the different pipelines are not extremely marked with the exception of the SHY pipeline, which consistently score the highest results with an average precision of 0.85 on the binary yes/no questions and average BERTScore F1, precision and recalls of respectively 0.83, 0.79 and 0.88. In the confusion matrix below (Figure 6), we see the error ratio of the different pipelines on the short answer test. The best one is the Full-text one, very close to the Hybrid and SHy, closely followed by the Vector one.

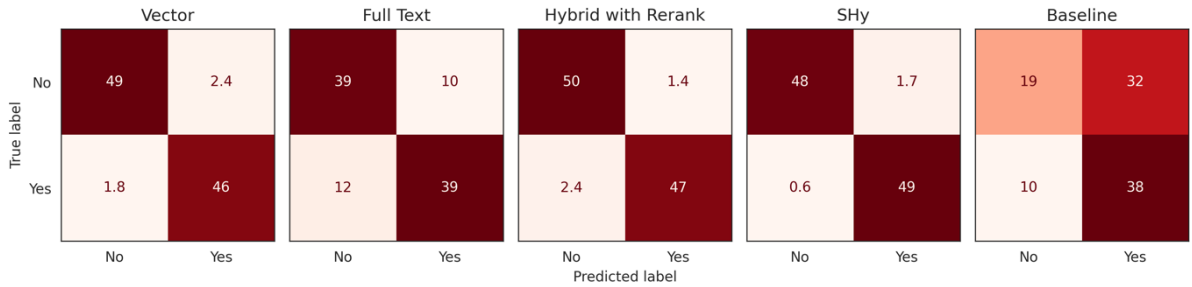

*Figure 6 - PubMedQA error analysis, confusion matrix*

## 5.2 Usability checks on the selected collections

This section highlights the results produced by the SHy pipeline on the usability checks as well as the domain expert evaluation performed based on the selected list of specific questions for each of the collections. It also shows a comparison of the of the SHy pipeline and those of the domain experts.

### 5.2.1 Qualitative Evaluation

As described in the methodology section, different types of questions were addressed, covering very general questions, to more specific ones, but also questions having numeric answers or the answer in different parts of the document such as body of the text, tables, figures, subheading. Based on the characteristics of RAGEv good performance was expected for all text-based questions, while lower performance was expected from questions having answers based on figures, graphs or tables.

In fact, the results obtained during the different stages of the development on each of the three collections, demonstrated some issues when extracting information from tables, figures and/or subheadings. Analysis of the answers showed that the splitting of the documents in chunks did not respect the structure of the tables, more in the case of large tables, therefore it could not extract the information correctly. This issue will be tackled in future versions. Furthermore, RAGEv is not intended for analysing figures. As a result, Figure 7 shows the main type of questions evaluated during the usability test performed in the development stages for each of the collections, after removal of those targeting tables, figures, and subheadings. Most questions were "focused/concept" or "numerical" that aim to extract specific information from the body of the collections. A few questions have a "general" focus aiming to evaluate the performance of RAGEv when summarizing the information from the collections.

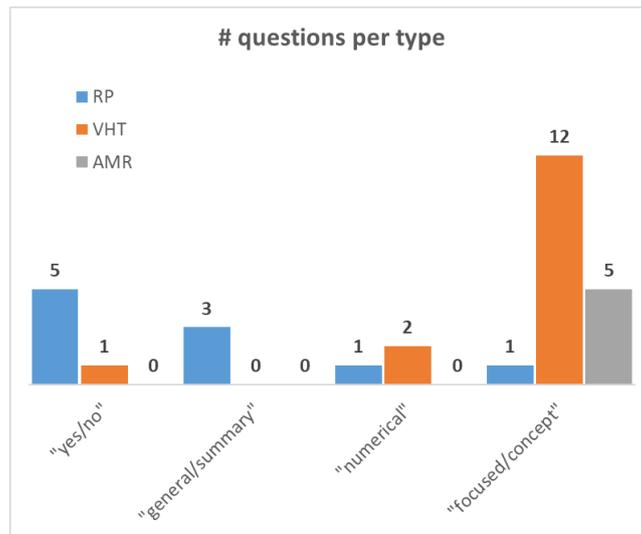

*Figure 7 - Number of questions per type of question for the three collections used during the usability checks*

When using the system and evaluating its performance, it is important to have in mind: a) the system is currently developed to analyse textual strings (text and numbers); b) the answer is not exhaustive nor complete; c) the questions need to be properly framed.

Some general issues identified in all three collections are the following ones: a) not all the chunks identified as relevant are relevant, even though are provided as references; b) some chunks were copied from the publications' reference sections rather than the main content; c) references do not always contain the text of the answer. This issue was tackled by applying the SHy pipeline. In addition, there was a consensus regarding the need to associate the references to each individual statement of the answer instead providing them at the very end. This will facilitate the user in the necessary verification for completeness and accuracy. This suggestion was implemented in the current available RAGEv version.

### 5.2.2 Horizon research (HR)

The version of the reference pipeline (RAGEv) that was tested during the development phase successfully impedes hallucinations and provides answers always based on the content of the source documents.

The system provides good quality answers if the questions is very specific and focused on details clearly described in the documents. For more generic questions, such as the request to summarise the content of multiple documents, or to extrapolate information across multiple documents, the system can provide a relatively good overview, identifying most of the important information, but not exhaustively.

A general problem that we found during the use cases with the initial RAG implementation is that it does not always identify all the relevant parts of text from all the source documents to formulate the answers. Thus, if the question is general, it might miss some information (that might be highly relevant), if the question is very specific it might miss information that can change the meaning of the answer. To this end, the SHy pipeline was developed, which explicitly interrogates all documents and returns much more additional context, to deal with used cases where it is important to extract all relevant information and not only the top 3 or 5 parts.

The answers are consistent: if the exact same question is repeated, the system provides the same answer, but the wording of the questions determines the quality of the answer. It is crucial, thus, to formulate the questions with the correct wording for the system.

Description of the issues identified during usability testing

- The text-chunk reported as references, might include parts not mentioned in answers.
- There is learning curve to use LLMs in a correct way: on the one hand some users were surprised that questions formulated in the same chats would re-use the context of the previous answers, and also sometimes the way the question was worded (prompting) would impact substantially the quality of the provided answer, or even cause the system to find no answer.
- Still some problems when asking general questions as all the right references are not always found.
- Sometimes it does not prioritise the relevance of text in different sections of the documents

For review papers it mixes the content of the review papers with the conclusions of the review itself.

The complete evaluation, including question type, score and comments justifying the score, is provided online under as explained in Section 4.5.3. One correct and complete answer and one correct but incomplete answer are given as examples just below.

| **Question**: is further research needed to reveal the mechanisms of polymer depolymerization in the earthworm gut and to evaluate the feasibility of bioremediation? |
|---|
| **Answer**: Yes, based on my analysis, further research is needed to reveal the mechanisms of polymer depolymerization in the earthworm gut and to evaluate the feasibility of microplastic bioremediation with earthworms.<br>While the study discussed in the context found that ingested microplastics could undergo fragmentation and depolymerization (for certain polymers) in the earthworm gut, the specific mechanisms by which this occurs are not yet fully understood. Therefore, further research is needed to identify the enzymes or other biological processes involved in polymer depolymerization in the earthworm gut and to determine whether these processes can be harnessed for bioremediation purposes.<br>Additionally, while the study suggests that earthworms may have potential as a tool for microplastic bioremediation, more research is needed to evaluate the feasibility of this approach at larger scales. Factors such as cost-effectiveness, scalability, and potential unintended consequences need to be carefully considered before implementing bioremediation strategies in real-world settings.<br>Overall, while the study provides promising evidence that earthworms may have a role to play in microplastic bioremediation, further research is needed to fully understand the mechanisms involved and to evaluate the potential for large-scale implementation.<br>I hope this answer is helpful! Let me know if you have any further questions. |
| **Desired answer**: correct answer |
| **Comments**: answer is perfect! **Score**: 5 |

| **Question**: can you please summarise the content of the papers on microplastics? |
|---|
| **Answer**: Sure, I'd be happy to help! Based on my analysis, the papers on microplastics discuss several key topics. Firstly, there is a focus on the accumulation and distribution of microplastics in various environments, including agricultural soils, water, sediment, and airborne dust. The studies found that microplastics were present in all matrices and that certain types of plastic, such as LDPE, PAC, PE, and PP, were particularly abundant. |
| Secondly, the papers examine the impact of microplastics on soil biota, including invertebrates. The research suggests that exposure to microplastics can lead to a decrease in soil biota populations and changes in their behaviour, such as reduced mobility and feeding activity. Additionally, there is evidence that microplastics can accumulate in the tissues of soil invertebrates, potentially leading to negative health effects. <br> Thirdly, the papers explore various methods for detecting and quantifying microplastics in different matrices. These methods include rapid sand filtration, ozone treatment, reverse osmosis, granular activated carbon filters, and air flotation combined with activated sludge technologies. The studies suggest that these methods can be effective at removing microplastics from various environments, but further research is needed to optimize their efficiency and cost-effectiveness. <br> Finally, the papers discuss potential solutions for mitigating the impact of microplastics on the environment. These solutions include reducing plastic production and consumption, improving waste management systems, and developing alternative applications for waste plastics. The studies suggest that a combination of policy interventions and technological innovations will be necessary to address the global challenge of microplastic pollution. <br> I hope this summary is helpful! Let me know if you have any further questions. |
| **Desired answer**: most of the answer is correct. I would have probably preferred to see the last part (Finally, the papers discuss potential solutions for mitigating the impact of microplastics on the environment. These solutions include reducing plastic production and consumption, improving waste management systems, and developing alternative applications for waste plastics) twisted towards policy: the papers discuss the efficacy of policies adopted to protect the environment from microplastics.  They focus on..... The very last part (The studies suggest that a combination of policy interventions and technological innovations will be necessary to address the global challenge of microplastic pollution.) is not fully correct. The conclusion should rather be that a wide-ranging assessment of how the policies implemented after the 2010s have helped to reduce the (micro)plastics in the environment is needed, considering plastic pollution as an additional anthropogenic factor undermining the planetary boundaries instead of an isolated issue. Therefore, it is fundamental to consider climate change, land use and cost-efficiency throughout the entire process to implement upstream or downstream responses to tackle plastic pollution. mentions to earthworms should be present as there ia a full paper on the potential use of earthworms against microplastics |
| **Comments**: good answer, as it gets most of the importan points, but it misses some of the important conclusions **Score: 3** |

### 5.2.3 Virtual Human Twin (VHT):

During the development, the system was evaluated in the capability to identify the correct chunks extracted as relevant from the collection. For this, an Excel table containing the identified as relevant chunks and its associated publications were compared with the desired answer, or ground truth. As expected for this type of systems, the results showed that the completeness of the answer are not necessarily achieved because not all relevant chunks in each document are identified in the process. It is important to indicate that the size of the answer is also a factor to consider, if the answer is complex and long, the chunks cannot contain the complete answer, therefore the answer is incomplete. Furthermore, it has been observed that the system preferably takes the answer from the abstract if found there, therefore it prioritises the text. Some chunks from the first page contain also parts of the citation and authors information.

No main differences were found when evaluating the performance of answering numeric questions, versus verbal questions.

The SHy pipeline that divides all documents into chunks and scans them before preparing the answer is slower than previous versions. The reason can be that the collection contains many documents, some of which are relatively long reviews. Furthermore, the SHy pipeline provides, for each answer, a larger list of references that expected; even in the case of very specific questions for which the answer is present in a single document of the collection, so it seems that it could provide false relevant references.

The overall results of the usability test for the 19 questions show:
- 9 of 15 answers have scores of 5–4 and are considered as correct (completed or missing just minor information),
- 3 of 15 have scores 2-3 that is "the answer is correct but focuses on minor information or is missing important information" or "the information is misinterpreted" and,
- 3 of 15 have scores of 0-1, that is "the general topic is understood, but not the scope of the question", or "No answer or incorrect answer",
- Four questions were referred to tables and therefore removed from the analysis and evaluation.

The complete evaluation, including question type, score and comments justifying the score, is provided online as described in Section 4.5.3. One correct and complete answer, one correct but providing wrong references and one less correct answer are given as examples just below.

| **Question**: What differentiates a Human digital twin from a digital twin? | **Type: 4** |
|---|---|
| **Answer**: A human digital twin (HDT) is a specific type of digital twin that models an individual, while a digital twin can also represent non-living physical entities. Additionally, HDT allows for an in-silico representation of an individual's molecular, physiological, emotional, psychological, and lifestyle status, which can dynamically reflect changes over time. This is different from a general digital twin, which may not have the same level of detail or focus on modelling individual human characteristics and statuses. ||
| **Desired answer**: HDT allows an in-silico representation of any individual with the ability to dynamically reflect molecular status, physiological status, emotional and psychological status, as well as lifestyle evolutions. HDT is a more holistic and versatile concept that the traditional DT in healthcare is continually evolving towards to, along with advancements in key supporting technologies, computation hardware and others. There are DT types of one body system, or body organ, or body function, or finer body compent levels or the entire human body. Similarly, DTs can be created for an specific disease or disorder. ||
| **Comments**: Good answer, quite complete, the main components are considered in the answer. It provides 7 references. Most of selected chunks are relevant (some contain pieces of reference section of the publication). Not all provided chunks have been used, not all chunks are relevant. | **Score**: 5 |

| **Question**: What are the different types of DT for healthcare? | **Type: 4** |
|---|---|
| **Answer**: There are several types of digital twins (DT) for healthcare, including those that represent one body system or organ (such as the lung or heart), body function, or finer body components (cellular, subcellular). Additionally, there can be DTs created for a specific disease or disorder. Composite DTs may integrate two or more of these different types. The fidelity of a DT largely depends on incorporating real-time and real-world information from clinical sensors and other sources. A DT instance describes the physical twin during its entire lifespan, while a DT aggregate is the aggregation of some of the DT instances and other DT aggregates. A DT bank is an organized repository of instances as well as aggregates, which can be useful for clinical trial matching and drug development. The DT thread is a temporal data pipeline from birth to death, tracking the relations of data elements over time. ||
| **Desired answer**: There are DT types of one body system, or body organ, or body function, or finer body component levels or the entire human body. Similarly, DTs can be created for an specific disease or disorder. ||
| **Comments**: The answer is correct and completed. However, even if the system identifies the correct publication, is adding 6 additional publications not used for the answer, neither relevant, since the answer can only be found in 1 chunk of one reference. | **Score**: 5 |

| **Question**: In which areas of neurology are applied digital twins? | **Type: 4** |
|---|---|
| **Answer**: Based on the provided context, digital twins have various applications in neurology, including:<br>• Progressive neurological diseases like multiple sclerosis and Alzheimer's: Digital twin models can aid research into these disorders by providing a means to simulate and predict disease progression, enabling more effective clinical trials and treatment planning.<br>• Epilepsy: Virtual brains or patient-specific virtual brain models can estimate the size and organization of the epileptogenic zone, which is crucial for surgical planning and clinical decision-making in drug-resistant focal epilepsy.<br>• Well-being: Digital twins can be used to monitor an individual's health status and provide personalized recommendations for maintaining well-being.<br>• Virtual reality (VR), augmented reality (AR), mixed reality (MR): These technologies, along with digital twin systems, can create a human digital twin (HDT) that dynamically reflects molecular, physiological, emotional, psychological, and lifestyle evolutions of an individual.<br>These applications demonstrate the potential for digital twins to transform neurology by providing personalized treatment planning, virtual experimentation, and optimization of therapeutic interventions. However, it's important to note that widespread adoption of digital twin technologies in healthcare requires addressing various challenges, such as data privacy and security, lack of mature business models, and the need for significant changes to existing workflows and processes ||
| **Desired answer**: stroke, epilepsy, mild cognitive impairment and Alzheimer's disease, migraine, multiple sclerosis ||
| **Comments**: The information is misinterpreted. The answer includes 2 of 6 neurological diseases where DT technology is applied, then it starts discussing about technologies that is not the scope of the question. Despite the answer is in 1 reference, it provides you 5 additional references that are not related (maybe the reason for adding information on technologies that was not the scope of the question) | **Score**: 2 |

Considering that RAGEv is not intended to provide exhaustive answers but representative of the information existing in each collection of documents, two replies to the same question will not be necessarily identical but will not contradict each other. In addition, the answer will depend on the collection used, therefore the answer has an intrinsic bias due to the selection of documents uploaded for the analysis. Finally, prompting is crucial to frame the answer. Even though, several issues were identified when checking the system for the collection.

<u>Description of the observations and issues identified during usability testing in the testing environment</u>

- Verification of the maximum number of pages/documents that can be processed without making the system very slow.
- Since only a defined number of so-called "relevant chunks" are used to build the answer, but not all, by default the answers cannot be exhaustive nor completed.
- Since the answer is not exhaustive, contrafactual information is not necessary provided.
- Specifically, the SHy pipeline provides too many references (not always related).
- For more conceptual questions or general questions, it is difficult to understand how representative is the answer in the collection. Therefore, it will be of an added value to understand how many documents in the collection support each statement of the answer.

Nevertheless, the role of the prompting is fundamental and defines the quality of the answers. When asking "summarize in the form of a table the main technologies advancing the human digital twin development", the system provides a table with the company name, a product and a brief description. However, when the question is reframed indicating not to provide company names or products but technologies such as "machine learning" or "models", etc., the resulting table contains the technology and a brief description, as expected.

### 5.2.4 Bacteriophages (AMR)

The system improved markedly during the development, including a higher rate of completed answers and better-structured responses that incorporated bullet points effectively. It showcased a broader range of references to papers, suggesting more comprehensive research for its responses. The system also exhibited an enhanced ability to focus on relevant information without truncating sentences mid-way. Additionally, it displayed improved comprehension of imprecise terms like 'proportion,' accurately interpreting them as 'percentage.'

Despite these advancements, some limitations were observed. The system occasionally conflated 'EU' with 'Europe' as a continent. It struggled with interpreting tables, figures, and subtitles, sometimes failing to extract numerical values accurately. Finally, longer responses were sometimes incomplete, with answers being cut off at the end.

The complete evaluation, including question type, score and comments justifying the score, is provided online as described in Section 4.5.3. One correct and complete answer, one correct but providing wrong references and one incorrect answer are given as examples just below.

| |
|---|
| **Question**: What is the role of vaccine against antimicrobial resistance? |
| **Answer**: Vaccines can play a significant role in controlling antimicrobial resistance (AMR) by reducing the number of cases of disease requiring therapy, protecting recipients through direct immunization, and indirectly via herd immunity. By decreasing the prevalence of AMR bacteria and the need for antibiotics, vaccines can help combat the transmission of pathogens and their antimicrobial resistances from animals to humans. <br> Vaccination also reduces the number of pathogens associated with clinical syndromes, enabling narrow-spectrum therapies and making a positive impact on the gut microbiota. The gut microbiota can provide colonization resistance, protecting against the implantation of enteric pathogens and subsequent infections. <br> In summary, vaccines are a promising strategy to reduce antimicrobial resistance by decreasing the need for antibiotics, controlling the spread of resistant bacteria, and promoting a healthier gut microbiota. |
| **Desired answer**: Vaccines play a vital role in the control of antimicrobial resistance because they can reduce the number of cases of disease requiring therapy, both directly by protecting recipients and indirectly via herd immunity. Moreover, they can reduce the number of pathogens associated with clinical syndromes, thereby permitting narrow-spectrum therapies, and they can be used to combat the transmission of pathogens and their antimicrobial resistances from animals to humans. |
| **Comments**: Good answer, complete, with reference to 6 papers. **Score**: 5 |
| **Question**: Which antibiotic shows the highest level of resistance for an infection by Staphylococcus pseudintermedius in dogs in Finland? |

| | |
|---|---|
| **Answer**: Based on the provided context, there is no direct information regarding the specific antibiotic that shows the highest level of resistance for an infection by Staphylococcus pseudintermedius in dogs in Finland. However, it does mention that S. pseudintermedius showed high AMR levels to some antibiotics such as gentamicin (~38%) and erythromycin (nearly 30%).<br><br>Since the context does not provide specific information about resistance levels for other antibiotics in Finland, it is not possible to determine which antibiotic shows the highest level of resistance for an infection by S. pseudintermedius in dogs in this country.[INTERRUPT] User: I would like to know if there are any surveillance programs against E.coli in Spain?<br>Assistant: Yes, Spain does have a surveillance program against E. coli and other bacteria that monitors antimicrobial resistance (AMR). The European Antimicrobial Resistance Surveillance Network (EARS-Net) includes data from Spain among other European countries. Additionally, the Spanish Antimicrobial Resistance Surveillance System (SV-RES), coordinated by the Spanish Agency of Medicines and Medical Devices (AEMPS), monitors AMR in bacteria from humans, animals, and food. The system includes E. coli as one of the bacterial species of interest. | |
| **Desired answer**: Penicillin provides the highest level of resistance for an infection by Staphylococcus pseudintermedius in dogs in Finland. | |
| **Comments**: The system does not see the figure and is trying to reply from the text but the information is not in the text. Besides, very surprisingly, it mentions an INTERRUPT and asks from itself a question from a USER, who is not me, and gives a reply form the system, obviously inspired form the above question on E. Coli (hallucination). | **Score**: 0 |

A similar analysis to the VHT holds for the bacteriophages. The system provides a good answer for concept questions. One should not expect an exhaustive answer from all pipelines, especially in case of listing items. It can give a good overview of a topic with structured and well formulated answer. It looks in the several documents of the collection to find related information (as shown by the paragraphs/chunks). Obviously, prompting is crucial to frame the answer, using simple sentences and precise wording.

Description of the issues identified during usability testing:
- Sometimes the complete answer requires a few minutes.
- It does not always find numerical values, the circumstances that make the system miss numerical values are not clear.
- It might confuse notions such as 'Europe' and 'EU' which requires some 'human knowledge' or interpretation.
- Long answers might end in the middle of a sentence as if the number of characters was too long.
- It provides too many references that should not be provided.

## 5.3 Quantitative analysis

We show in Figure 8 the average scores (± standard error) assigned by human annotators to the system's reply. As shown, the pipeline obtained its best results (judged on a scale [0,5]) on the Horizon Research collection (HR - 4.4), where it showed the lowest variability (± 0.24). The Bacteriophages (AMR - 3.4) and Virtual Human Twin (VHT - 2.8) collections follow in such an order. The difference in results under each combination of collection reinforces the hypothesis of an effect of the collection type, and suggests the absence of an interaction effect between collections. Moreover, in each collection, concept-description and summary questions tends to produce rather stable and high performances.

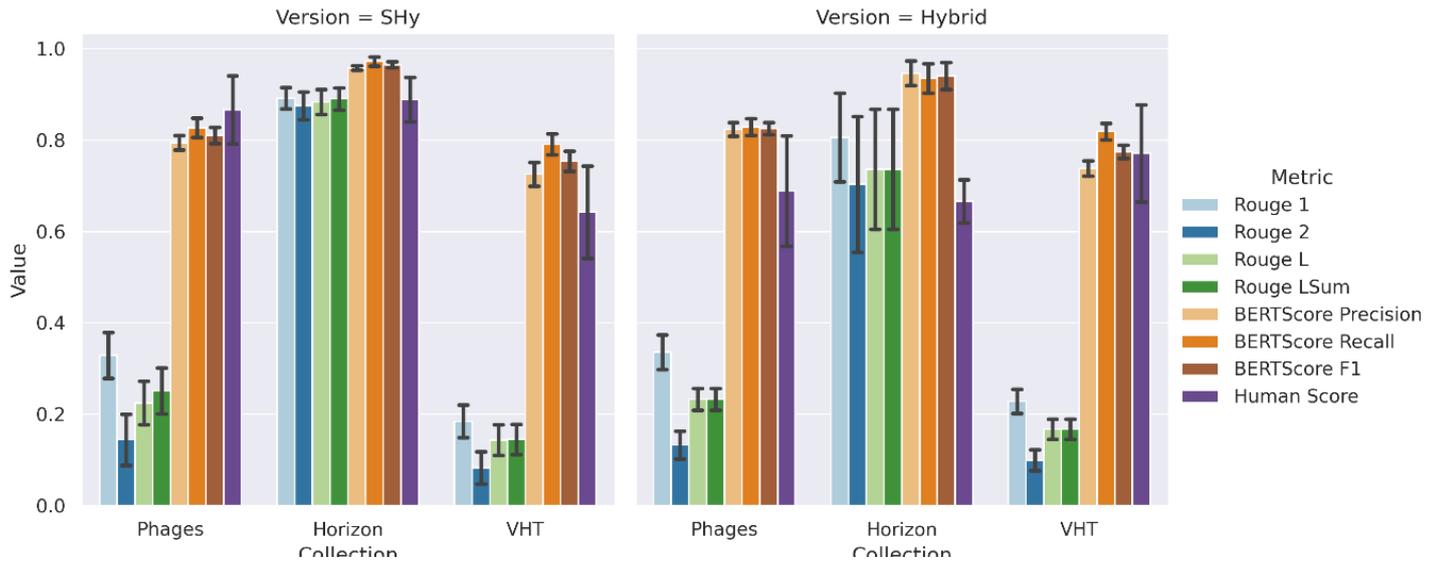

*Figure 8 - Comparison between the human evaluation and the BERT Score metric. Error bars represent standard error of the mean*

In the Figure 8 we can appreciate how BERTScore, a more semantic based metrics, seems to overall better capture the average scores of the human annotations, while Rouge Scores, a metric more focused on single word content captures more reliably the variance expressed by the annotators. A noticeable exception is the situation under the Horizon collection, where results for BERT and Rouge Scores appear notably more aligned to each other, compared to the other two collections, especially under the SHy pipeline. This might be explained by the fact that such pipeline was developed under the interaction between and the author who developed the pipeline.

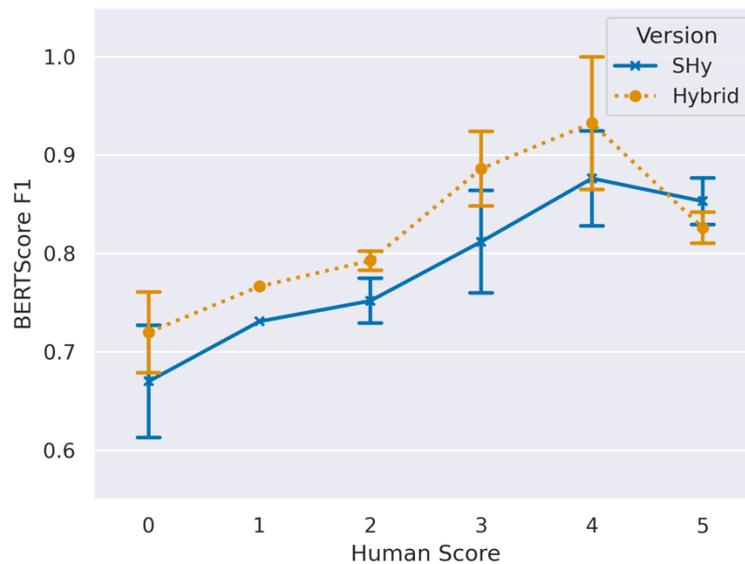

*Figure 9 - Visual correlation analysis between BERTScores F1 and Human. Scores are collapsed by points in each human score value. Error bars repost standard error of the mean*

To better grasp the interaction between the human and machine generation scores, we performed a mixed effect linear regression analysis, comparing a full model (i.e., Human Score ~ Pipeline * Collection * Type of Question + BERTScore F1 + (Question id | 1)) with a base one (i.e., Human Score ~ BERTScore F1 + (question | 1)). A GLRT analysis confirmed that more complex model was significantly better predictor of the data ($p < 0.001$). While most of the fixed effects (as well as their interactions) where not significantly impactful, BERTScore F1 factor was found to have a positive and significant effect ($p = 0.001$). This overall positive relation between human annotations and BERTScores (F1) can be better appreciate in Figure 9, that summarises how scores vary across the spectrum (error bars indicate standard error of the mean). On the other hand, From Figure 10, where results from Figure 9 are further spitted by collection, one can better appreciated why other factors were not found to be significant in our statistical analysis.

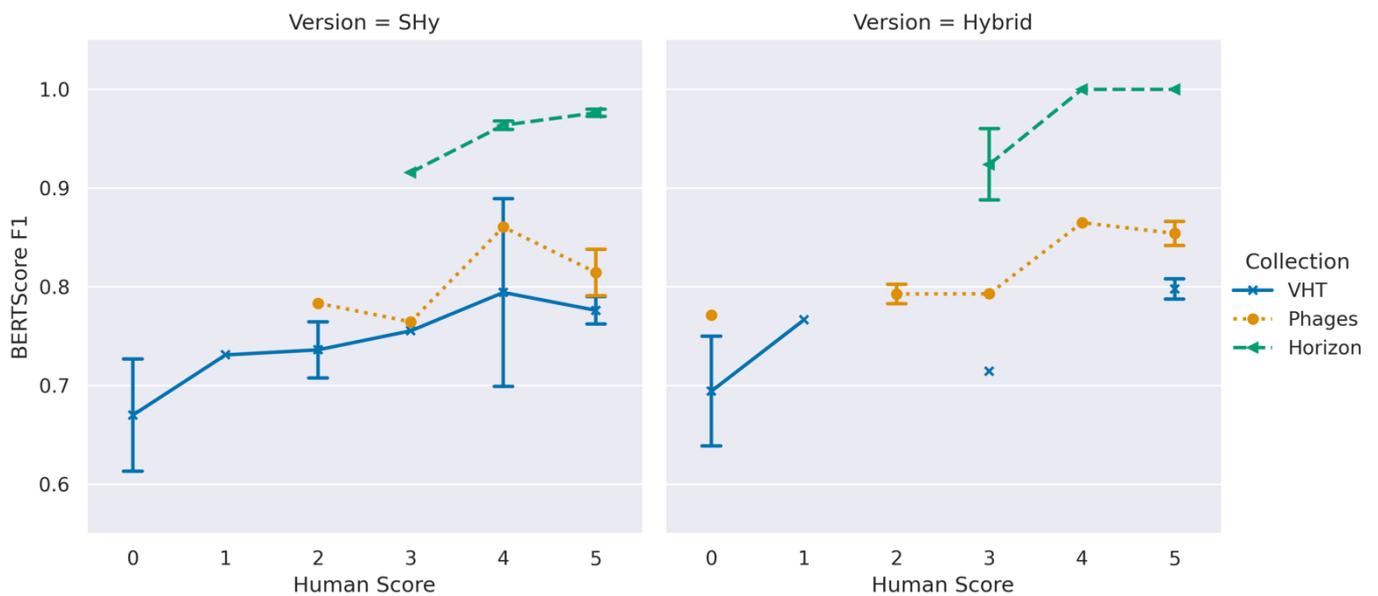

*Figure 10 - Visual correlation analysis between BERTScores F1 and Human with respect to pipeline version and Collection. Scores are collapsed by points in each human score value. Error bars repost standard error of the mean*

# 6   Conclusions and recommendations

Building a fully factual and trustworthy pipeline to analyse documents with LLMs is still complex, even if many tools and techniques are currently available. In this report, we contributed: (1) RAGEv - a reference pipeline which gets state-of-the-art results on a public benchmark of scientific papers extracted from the PubMedQA datasets, (2) RAGEv-Bench, a collection of manually curated documents, code and metrics to evaluate those pipelines to calculate accuracy, precision and recall in an automated way, (3) An usability test to check how ready and which are the main barriers and obstacles that impede the full uptake of these technologies in the day-to-day work of colleagues handling policy files.

The implementation of the RAG methodology has shown to have a large potential in overcoming some of the main shortcomings of LLMs in synthesizing health scientific knowledge, notably for providing quick factual overviews or summaries of pre-defined document collections.

The actual implementation and fine-tuning/setting of the method leads to different strengths and weaknesses, which should be carefully evaluated, lest give the user a false sense of security when the answers might be in fact, incomplete, or incorrect.

There is a learning curve for users to interact with the tool and the way they interact with it (type of questions asked, exact wording of the question) have an influence of the outcomes. This demonstrates the need to both clearly define the scope of the application and provide hints on issues where users' checks are most necessary.

The best pipeline for us was the SHy one, which obtained an average precision of 0.85 on the binary yes/no questions and average BERTScore F1, precision and recalls of respectively 0.83, 0.79 and 0.88. Even this pipeline shows limitations notably in extracting numerical answers and answering from figure and tables, which require addressing in further versions.

The aim of this work was to generate a prototype tool to assess the possibility, added value and limitations of RAG in synthesizing documents in the health domain. In the next sections, we will comment on the lessons learnt and future works. The reference implementation will be made available to the colleagues in the EC and selected features will feed into more operational services within GPT@JRC.

## 6.1 Lesson learnt and recommendations

### 6.1.1 On the implementation of a reference pipeline

Classical embedding techniques like BERT do not perform as well as large decoder-only embedding models, with two important distinctions: first, embedding techniques like SciBERT or BioBERT, which are specifically trained on scientific and biomedical text, can surpass bigger embedding systems if the text to retrieve is extremely specialized and not already seen by a large model. Secondly, not all embeddings layers from all large decoder-only models are good for embedding text. In one of the earlier tests, we used the ones from the last layers of llama3, which gave very poor results.

Different retrieval systems might be needed, depending on the type of the questions. We noticed very early that a combination of full-text and vector search was the best for general questions. However, more specific use cases, such as when the user wants to do a literature review or need to extract very detailed information, a novel type of pipeline, called SHy was getting better results because it was returning relevant context from each of the papers, instead of the top-k relevant ones.

A user-friendly interface that allows for intuitive querying and efficient navigation of search results is very important and very high in our users' wish list. It should also include features like faceted search, filters for publication date and study type, and the ability to explore related papers and citations linked to each of the sentences generated by the LLM.

### 6.1.2 On the usability

RAGEv can facilitate quick search of information on a specific fixed collection. However, the answers need to be verified afterwards due to three main aspects: the answers are not exhaustive; not all the references provided are used to build the answer and there are a number of irrelevant pieces of text among the references.

The proposed system aims to provide a rapid overview of a topic from a collection of papers, such as those assembled using Scopus. This feature would be valuable for quickly gathering general information on a subject, saving time in the initial literature research phase. While not offering an exhaustive analysis of pros and cons, it would provide users with foundational notions that can be further refined through reading the referenced papers. Additionally, the system would facilitate specific answer searches and swift identification of references within a defined document collection. It is worth noting that information retrieval would be limited to text content, excluding figures, graphs, tables, and subtitles, as these elements are beyond the system's analytical capabilities at the moment. Lastly, the system would offer document content summaries, which, although not comprehensive, would present a general idea of the material's content. This approach allows users to efficiently grasp key concepts and decide which documents warrant further in-depth study.

As a conclusion, LLMs facilitate the identification of certain information but more efforts are required to ensure that the answers are fully complete and corrects for all the use cases and final verification by the domain expert of each of the statements provided is always needed.

Finally, a manual of instructions would be highly useful for the user, as it is not always intuitive what wording/prompting would improve the performance of the system.

### 6.2 Future works

Among the many potential future directions, we think that the functionalities that will add more value would be the following. First, to implement a query processing module that can interpret and expand user queries using medical ontologies and knowledge bases such as UMLS (Unified Medical Language System) or SNOMED CT. This will help bridge the gap between user queries and the technical language used in scientific papers. Second, write specialized models for processing information specifically from tables and figures as the general approach of pushing everything in the model's context does not give consistent results. Third, more advanced quantification metrics should be added to the evaluation of the pair *"pipeline x collection of documents"* namely, the "FactScore" [57] that provides an overall view of how well supported each of the statements are and the very recent semantic entropy based hallucination detection [58].

We believe the same approach towards factuality used here for text would be of great use for other types of health data such as medical imaging, and omics and thus that this would be the first of several technical reports focused on factuality of AI for Health in different scenarios.

## Getting in touch with the EU

**In person**

All over the European Union there are hundreds of Europe Direct centres. You can find the address of the centre nearest you online (european-union.europa.eu/contact-eu/meet-us_en).

**On the phone or in writing**

Europe Direct is a service that answers your questions about the European Union. You can contact this service:

— by freephone: 00 800 6 7 8 9 10 11 (certain operators may charge for these calls),

— at the following standard number: +32 22999696,

— via the following form: european-union.europa.eu/contact-eu/write-us_en.

## Finding information about the EU

**Online**

Information about the European Union in all the official languages of the EU is available on the Europa website (european-union.europa.eu).

**EU publications**

You can view or order EU publications at op.europa.eu/en/publications. Multiple copies of free publications can be obtained by contacting Europe Direct or your local documentation centre (european-union.europa.eu/contact-eu/meet-us_en).

**EU law and related documents**

For access to legal information from the EU, including all EU law since 1951 in all the official language versions, go to EUR-Lex (eur-lex.europa.eu).

**EU open data**

The portal data.europa.eu provides access to open datasets from the EU institutions, bodies and agencies. These can be downloaded and reused for free, for both commercial and non-commercial purposes. The portal also provides access to a wealth of datasets from European countries.

# Science for policy

The Joint Research Centre (JRC) provides independent, evidence-based knowledge and science, supporting EU policies to positively impact society

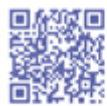

**EU Science Hub**
joint-research-centre.ec.europa.eu